\documentclass[aps,prb,twocolumn,groupedaddress,floatfix,showpacs]{revtex4}

\usepackage{graphics}
\usepackage{epsfig}
\usepackage[latin1]{inputenc}
\usepackage{subfigure}
\usepackage{dcolumn}
\usepackage{bm}
\usepackage{overpic}

\begin{document}

\title{Solar induced growth of silver nanocrystals}

\author{Annett Th\o gersen$^{1,2}$}

\affiliation{$^1$Department of Solar Energy, Institute for Energy Technology, Instituttveien 18, 2007 Kjeller, Norway}
\address{$^2$SINTEF Materials and Chemistry, P.O.Box 124 Blindern, 0314 Oslo, Norway}

\author{Georg Muntingh}
\affiliation{Centre of Mathematics for Applications, University of Oslo, P.O.Box 1053, Blindern, 0316 Oslo, Norway}

\date{\today}

\begin{abstract}
\noindent The effect of solar irradiation on plasmonic silver nanocrystals has been investigated using Transmission Electron Microscopy and size distribution analysis, in the context of solar cell applications for light harvesting. Starting from an initial collection of spherical nanocrystals on a carbon film whose sizes are log-normally distributed, solar irradiation causes the nanocrystals to grow, with one particle reaching a diameter of 638 nm after four hours of irradiation. In addition some of the larger particles lose their spherical shape. The average nanocrystal diameter was found to grow as predicted by the Ostwald ripening model, taking into account the range of area fractions of the samples. The size distribution stays approximately log-normal and does not reach one of the steady-state size distributions predicted by the Ostwald ripening model. This might be explained by the system being in a transient state.
\end{abstract}

\maketitle

\section{Introduction}
\noindent Metallic nanocrystals are desirable for their use in solar cell applications, where they can be used to increase efficiency by means of the localized surface plasmon resonance. This collective oscillation of the valence electrons results in a local electromagnetic field at the nanocrystal surface and in wavelength selective photon absorption and scattering \cite{intro:1}. By dispersing metal nanocrystals on the surface of an optically thin solar cell, the localized surface plasmon resonance will scatter the light further into the solar cell, thereby increasing the absorption and thus the efficiency.

Silver (Ag) nanocrystals are particularly popular, being relatively cheap and easy to produce by a variety of processes. In addition, they do not need to be deposited and heat-treated, which can be damaging for the solar cells. Using Ag nanocrystals, Pillai \cite{intro:31} and Beck et al. \cite{intro:beck} demonstrated a significant increase in absorption in wafer-based silicon solar cells. However, nanocrystals used in solar cell applications will be exposed to elevated temperatures, which can make them unstable. A solid understanding of the underlying growth and coalescence mechanisms is therefore of vital importance.

The Ag nanocrystals in this work have been made by a chemical synthesis of metallic nanocrystals, using chemical reduction of silver salts by sodium borohydride. This has shown to be a simple, economical, and popular method that can be applied in large scale as required for industrial applications\cite{intro:2,jack:emrs}. The surface plasmon resonance depends on the shape of the nanocrystal surface, size, spatial arrangement, and configuration of the nanocrystals \cite{intro:4}. It is therefore very important to investigate these features in detail. We have previously studied the particle size distribution, structure \cite{annett:mrs} and optical properties of colloidal Ag nanocrystals \cite{annett:ag1, jack:emrs}. These nanocrystals, however, must remain stable during solar irradiation and heat. In this paper we study the effect of solar irradiation on Ag nanocrystals on a carbon film, and investigate the growth rate and particle distribution with solar irradiation time.

\section{Experimental}
\noindent The Ag nanocrystals were made by a wet chemical reduction synthesis mixing a silver nitrate solution (AgNO$_3$, 99.9\% purity, Qualigens Fine Chemicals) with a highly reducing solution of sodium borohydride (NaBH$_4$, 95\% purity, Merck) without further purification. All equipment was thoroughly cleaned by soaking in ethanol and washed with distilled water. A 0.6 mM AgNO$_3$ precursor solution was prepared using deionized H$_2$O, together with a 1.2 mM aqueous NaBH$_4$ solution. In order to prevent agglomeration, the solutions were kept ice cold and the mixture was stirred continuously. Ag nanocrystals were formed by adding the aqueous AgNO$_3$ drop by drop to the highly reducing solution of NaBH$_4$. Variations in particle size and distribution were acquired by varying the amount of AgNO$_3$ (aq) relative to NaBH$_4$ (aq) in the final mixture. The samples studied in this work have a AgNO$_3$/NaBH$_4$ molar ratio of 2/25.

Six Transmission Electron Microscopy (TEM) samples were prepared by adding a drop of the final solution on a holy carbon film supported by a Cu grid. In order to observe light induced effects, five of the TEM samples were placed under a Xenon lamp, irradiating the wafers with an intensity of 100 mW/cm$^2$ (1 Sun). While the samples started off at room temperature, the temperature slowly increased from 55.7$^\circ$C after 15 min.~to 64.3$^\circ$C after 240 min. The lamp was placed 35 cm from the TEM sample grids. The remaining sample, which we shall refer to as sample 1, was kept aside. One by one the other samples are then taken away from the Xenon lamp, sample 2 after 15 min., sample 3 after 30 min., sample 4 after 60 min., sample 5 after 120 min., and finally sample 6 after 240 min. The Ag nanocrystals were then observed by High-Resolution Transmission Electron Microscopy in a 200 keV JEOL 2010F microscope with a Gatan imaging filter and detector. 

\section{Theoretical}
\subsection{Ostwald ripening}
\noindent Ostwald ripening is a thermodynamically-driven spontaneous process, in which small particles dissolve and through diffusion redeposit onto larger particles in an effort to decrease the total free surface energy.

While first qualitatively described by Ostwald \cite{ostwald:2,ostwald:1}, the theory made a major leap forward when Lifshitz and Slyozov \cite{LW:1} and Wagner \cite{wagner:LW} independently derived quantitative predictions about the distribution and evolution of the particle sizes. The LSW model, as it is now called, was derived under several assumptions, including:
\begin{enumerate}
\item The particles are spherical.
\item The distances between the particles are infinitely large compared to the size of the particles.
\end{enumerate}
In practice, however, particles are not always spherical, and infinite distances are physically unrealistic; the discrepancy is characterized in terms of the \emph{volume fraction} $\phi$ (or \emph{area fraction} in 2D), which is defined as the volumetric density of the particles. It is therefore no surprise that the size distribution predicted by the LSW model corresponds to virtually none of the size distributions encountered in experiments, which are generally broader and more symmetric \cite{Voorhees85}.

To address these shortcomings several extensions to positive volume fractions were proposed, including analytical calculations \cite{Hua:14} and numerical models \cite{Akaiwa:13, Akaiwa:18}, some extending to nonspherical particles \cite{fan:12}. For a comprehensive survey on the progress in Ostwald ripening theories, see the paper by Baldman \cite{baldman:19}.

One common feature of these extensions is the prediction that the average particle radius $\langle R \rangle$ follows the growth law
\[
\langle R \rangle^3 = \langle R_0 \rangle^3 + k_\phi t, 
\]
where $\langle R_0 \rangle$ is the average particle radius at $t = 0$, and the \emph{kinetic coefficient} $k_\phi$ depends on the area fraction $\phi$ of the sample. Although $k_\phi$ almost always increases with $\phi$, there are known situations where it decreases with $\phi$ instead \cite{Kim:17}.

Since some of our particles become moderately aspherical over time, they do not have a well-defined radius. We simplify our analysis by assuming the particles are spherical, and then computing their diameter $d$ from the measured cross-sectional area $A$ from the formula $A = \pi(d/2)^2$. As the diameter of a spherical particle is twice its radius, we expect the average diameter $\langle d\rangle$ to satisfy a growth law
\begin{equation}\label{eq:AverageGrowthLaw}
\langle d \rangle^3 = \langle d_0 \rangle^3 + 8k_\phi t.
\end{equation}
Numerical simulations by Fan et al. \cite{fan:12} suggest that
\begin{equation}\label{eq:KineticCoefficients}
k_{0.25} \approx 0.810,\  k_{0.5} \approx 2.548, \ k_{0.75} \approx 8.207,\ k_{0.9}\approx 24.487.
\end{equation}

\subsection{The log-normal distribution}
\noindent In our previous work on the growth of colloidal Ag nanocrystals, the particle sizes were shown to be log-normally distributed \cite{annett:ag1}. The discussion gave insight into the growth and structure of the formation process and provided a close fit with the experimental observations of the particle sizes.

The log-normal distribution with parameters $\mu$ and $\sigma$ has probability density function
\begin{equation}
\label{eq:1}
p(x; \mu, \sigma) = \frac{1}{x \sigma \sqrt{2 \pi}} \mathrm{e}^{ - \frac{ (\ln x- \mu )^2 }{2 \sigma^2} } ,\qquad x>0.
\end{equation}
The parameters $\mu$ and $\sigma$ coincide with the mean and standard deviation of the natural logarithm of the random variable $X$, which is by definition normally distributed. The random variable $X$ itself has mean
\begin{equation}\label{eq:mean}
m = \mathrm{e}^{\mu + \sigma^2/2}
\end{equation}
and standard deviation
\begin{equation}\label{eq:sd}
sd = \mathrm{e}^{\mu + \sigma^2/2} \sqrt{\mathrm{e}^{\sigma^2} - 1}.
\end{equation}

\subsection{Parameter estimation}
\noindent In our previous paper we chose to smoothen the overly erratic particle size densities by computing moving averages \cite{annett:ag1}. Since there is no clear choice for the size of the sampling window for the data in this paper, we instead choose to group the diameters into bins of each 1 nm in width.

The probability that the random variable $X$ assumes a value between $d-1/2$ and $d+1/2$ is
\[ P_{d;\mu,\sigma} := \int_{d-1/2}^{d+1/2} p(x; \mu, \sigma) \mathrm{d} x . \]
Suppose now that sample $i$ comprises $N_i$ particles of which $N_{i,d}$ are of size between $d - 1/2$ and $d + 1/2$. Then the log-likelihood is \cite{Raabe71}
\begin{equation}\label{eq:LogLikelihood}
\sum_d N_{i,d} \log(P_{d;\mu,\sigma}).
\end{equation}
The maximum likelihood estimators $\widehat{\mu}, \widehat{\sigma}$ of $\mu, \sigma$ are found by maximizing this expression numerically.

Plotting, for each sample, the mean nanocrystal diameter against the irradiation time suggests that there exists a relation (\ref{eq:AverageGrowthLaw}) between these two quantities. For a set of uncorrelated data with varying uncertainties, as is the case here, it is unreasonable to assume that every data point should be treated equally in estimating an underlying relation. Instead, a \emph{best linear unbiased estimator} is obtained by weighing each data point by the reciprocal of its variance \cite{Aitken35}. This \emph{weighted least square} method is well known, and was briefly described in a similar context in our previous work \cite{annett:ag1}.

\section{Results}
\noindent The size and shape of the particles has been studied with TEM, before and after solar irradiation. Initially the particles are spherical and small, with an average diameter of $\sim$6 nm. During the solar irradiation most of the particles grow, at the cost of the smaller particles. Some of the particles grow very large, the largest observed particle having a diameter of 638 nm after four hours of solar irradiation, and lose their spherical shape. Not all particles grow, however, as the sample irradiated for four hours still contains nanocrystals with the minimal observable diameter. Electron Energy Loss Spectroscopy revealed very little oxidation of the Ag nanocrystals. Only the outermost surface of the largest nanocrystals seems to be affected.

\begin{figure}
  \begin{center}
    \epsfig{figure=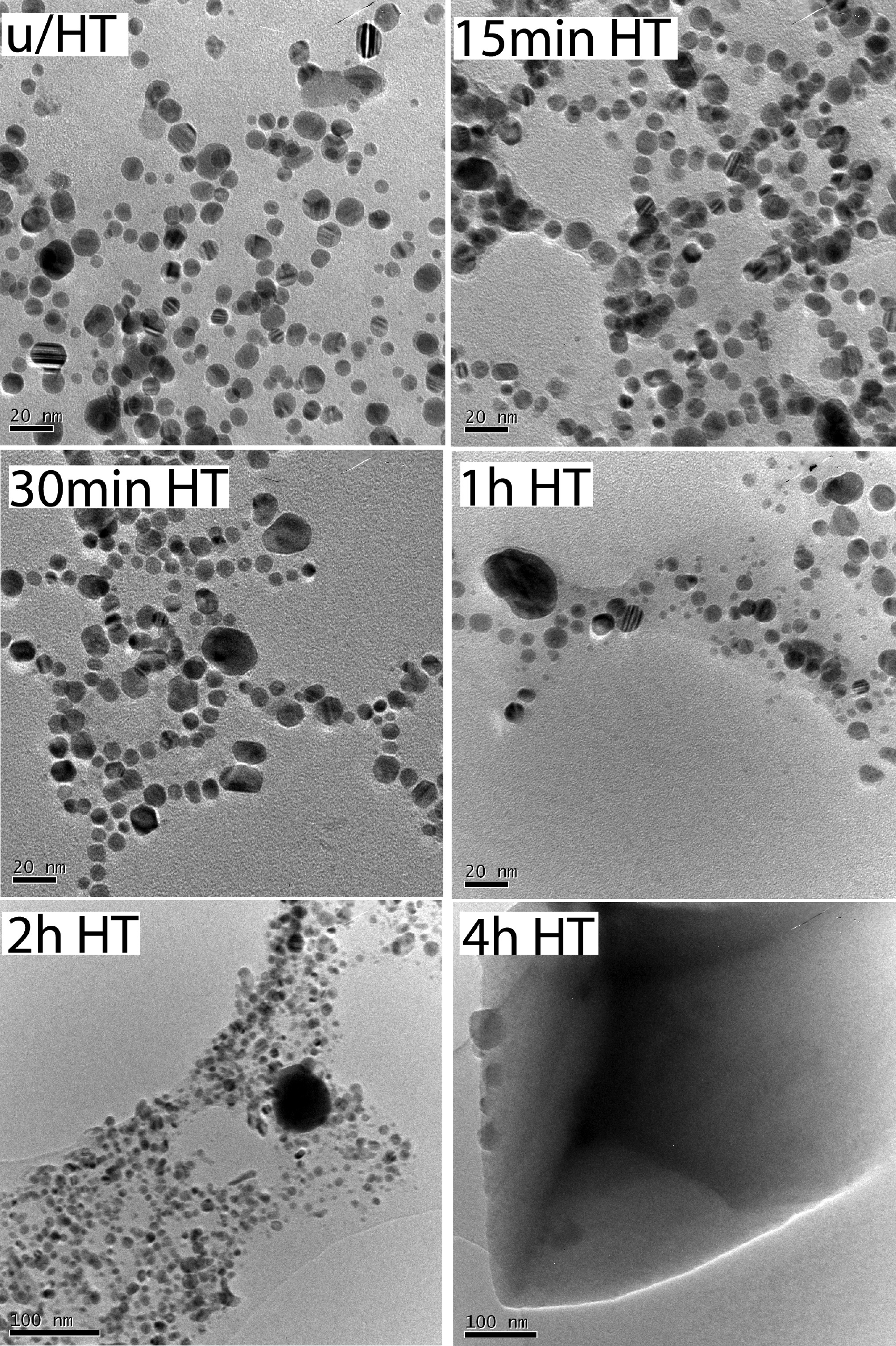,width=0.48\textwidth}
   \caption{TEM images of the Ag nanocrystals after solar irradiating for 0, 15, 30, 60, 120, and 240 minutes. Note that the images of samples 5 and 6 are taken at a different resolution.}
    \label{fig:TEM}
  \end{center}
\end{figure}

\subsection{Particle distribution}
\begin{figure}
\includegraphics[scale=0.65]{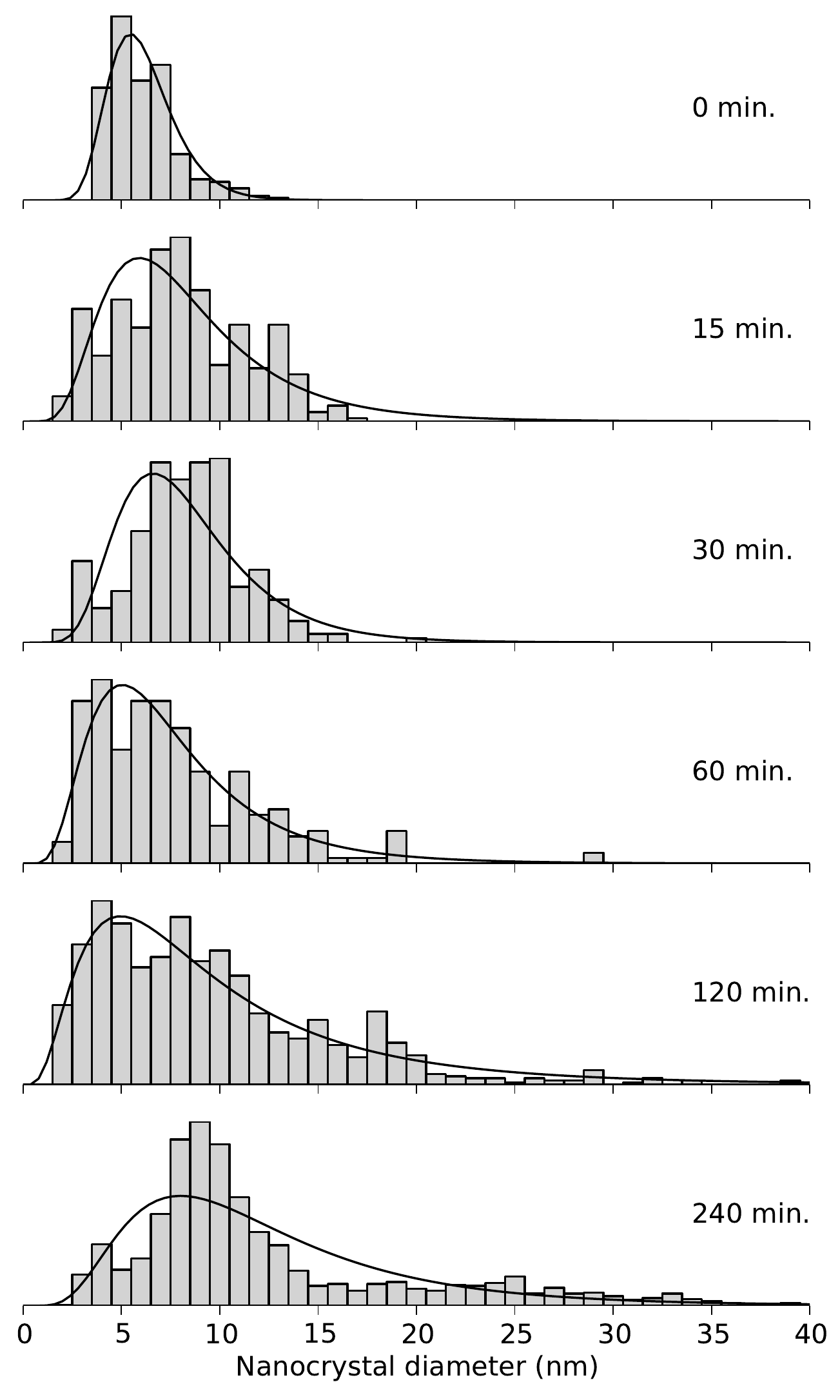}
\caption{The figure shows the relative number of Ag nanocrystal for each diameter, after solar irradiation for 0, 15, 30, 60, 120, and 240 minutes. Also drawn are fitted log-normal densities, whose parameters $\mu$ and $\sigma$ were estimated by maximizing the log-likelihood (\ref{eq:LogLikelihood}) numerically.}
\label{fig:EmpiricalDensities}
\end{figure}

\noindent For each sample at least six TEM images were taken and analyzed; see Figure~\ref{fig:TEM} for one image of each sample. The program ImageJ \cite{ImageJ} was used to compute the area fractions in the samples, and to accurately count the particles and estimate their diameters. The average area fraction in each sample varied from 0.240 to 0.463, while the number of counted nanocrystals varied from 257 to 1499. Since the bigger nanocrystals in sample 6 (partially) overlap many of the smaller nanocrystals, the nanocrystals in these TEM images were separated manually before applying ImageJ. A qualitative check was carried out by manually measuring the diameters for a smaller number of representative nanocrystals. Figure~\ref{fig:EmpiricalDensities} shows, for each sample, a histogram of the diameters, the bins being of 1 nm in width and centred at integral diameters.

Next, the maximum likelihood estimators $\widehat{\mu}, \widehat{\sigma}$ are found by maximizing (\ref{eq:LogLikelihood}) numerically, and the mean and standard deviation are computed from (\ref{eq:mean}) and (\ref{eq:sd}). These values are listed in Table \ref{tab:data}, and the estimated log-normal densities are plotted in Figure \ref{fig:EmpiricalDensities}.

\begin{table}
\begin{tabular*}{\columnwidth}{@{\extracolsep{\stretch{1}}}*{7}{r}@{}}
\hline
                 $i$ &     1 &     2 &     3 &     4 &      5 &      6\\ \hline
               $t_i$ &     0 &    15 &    30 &    60 &    120 &    240\\ 
               $N_i$ &   523 &   425 &   284 &   257 &    909 &   1499\\ 
   $\widehat{\mu}_i$ & 1.775 & 2.014 & 2.050 & 1.903 &  2.102 &  2.382\\ 
$\widehat{\sigma}_i$ & 0.279 & 0.487 & 0.401 & 0.535 &  0.708 & 0.547\\ 
               $m_i$ & 6.134 & 8.437 & 8.416 & 7.737 & 10.513 & 12.572\\ 
              $sd_i$ & 1.743 & 4.362 & 3.515 & 4.453 &  8.480 &  7.425\\ \hline
\end{tabular*}
\caption{For each of the samples $i = 1, 2, \ldots, 6$, the table lists the irradiation time $t_i$ in minutes, the number $N_i$ of counted nanocrystals, the parameters $\mu_i$ and $\sigma_i$ of the fitted log-normal distributions, and the corresponding expectation value $m_i$ and standard deviation $sd_i$.}
\label{tab:data}
\end{table}

\subsection{Particle growth}
\begin{figure}[t!]
  \begin{center}
    \includegraphics[scale=0.6]{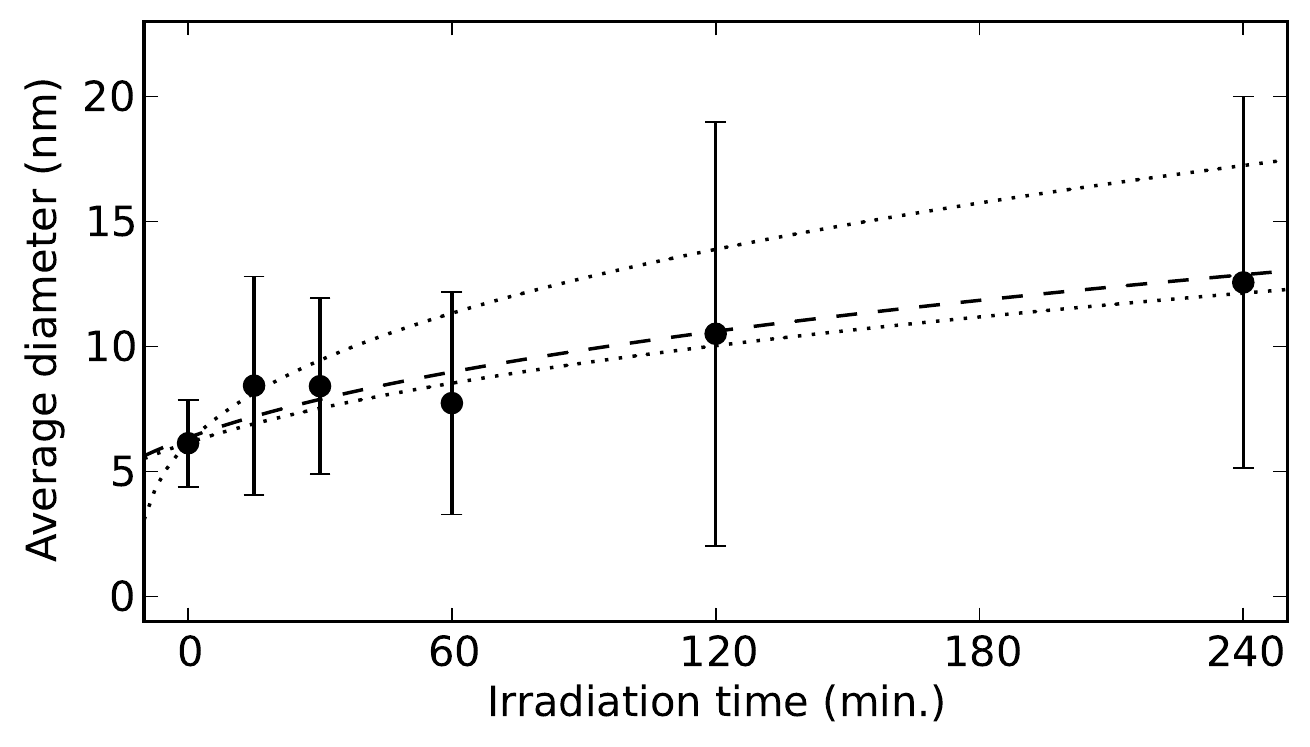}
   \caption{Average diameters against the irradiation times, with the error bar extending one standard deviation in each direction. The dashed curve is the growth law (\ref{eq:AverageGrowthLaw}) estimated by the weighted least square method. Drawn dotted are the predicted curves corresponding to the minimal and maximal area fraction.}
    \label{fig:AverageDiameter}
  \end{center}
\end{figure}

\noindent Table \ref{tab:data} suggests that the mean nanocrystal diameters $m_i$ grow with the irradiation time $t_i$. For a quantitative analysis we plot these quantities against each other in Figure~\ref{fig:AverageDiameter}, with error bounds the corresponding standard deviations $sd_i$. Drawn dotted are the curves (\ref{eq:AverageGrowthLaw}) corresponding to the minimal area fraction ($\approx 0.25$) and maximal area fraction ($\approx 0.5$), with $\langle d_0\rangle \approx 6.152$ the empirical mean diameter of sample 1 and $k_\phi$ as in (\ref{eq:KineticCoefficients}) predicted by the numerical simulation by Fan et al. \cite{fan:12}. Applying the weighted least square method, with weights the reciprocals of the variances $sd^2_i$, yields a fit of (\ref{eq:AverageGrowthLaw}) with 
$\langle d_0\rangle = 6.429$ and $k_\phi = 0.881$.
The corresponding curve is plotted (dashed) in the same figure.

\section{Discussion}

\subsection{Particle distribution}
\noindent As shown in a previous work \cite{annett:ag1}, the log-normal distribution provides a close fit for the diameters of the nanocrystals in sample 1. After irradiating the samples for various times, the log-normal distribution continues to be a good fit for the diameter distributions for samples 2 -- 5. In sample 6 the log-normal distribution no longer provides a good fit, and the steady-state size distributions predicted by the Ostwald ripening model do not fit either. However, the peak is becoming more pronounced, and after scaling each empirical density by the average diameter, the tail seems to be moving inwards. These features are consistent with the system being in a transient state \cite{Snyder.Alkemper.Voorhees}.

The evolution of the particle size distribution under the Ostwald ripening is notoriously slow, and the time required to reach a steady state generally depends on the initial distribution \cite{Chen.Voorhees.93} and the area fraction \cite{Snyder.Alkemper.Voorhees}. Compared to sample 1, the average diameter in sample 6 has changed by a factor of $m_6/m_1 = 12.572/6.134 \approx 2.05$. This is typically too little to reach the steady state \cite{Snyder.Alkemper.Voorhees}, another indication that the observed distribution is in a transient regime. More data, in particular data for longer irradiation times, are needed to confirm this phenomenon.

\subsection{Particle growth}
\noindent With longer solar irradiation the particles grow and become aspherical. Such aspherical particles are common for higher area fractions \cite{fan:12}. The growth of the average diameter of the nanocrystals fits the growth law (\ref{eq:AverageGrowthLaw}) predicted by the Ostwald ripening model. Because the area fraction $\phi$ varies from sample to sample, the kinetic coefficient $k_\phi$ takes on different values as well. As can be seen in Figure \ref{fig:AverageDiameter}, however, the least square estimate fits well between the corresponding range of kinetic coefficients predicted by Fan et al. \cite{fan:12}.

As seen in our previous work \cite{annett:ag1}, the increase in particle size will yield an increase in the wavelength of maximum scattering. This will increase the efficiency of the device initially, since the nanocrystals investigated in that paper have a maximum scattering peak at relatively low wavelengths (e.g. $\sim 480$ nm in silicon nitride), whereas the optimal efficiency for the solar spectrum is attained above 600 nm, for instance at 700 nm. Since the device would be unstable, however, its efficiency would change with time and would eventually decrease again.

\section{Conclusion}
\noindent The effect of solar irradiation on Ag nanocrystals has been investigated by TEM and size distribution analysis. Solar irradiation causes the nanocrystals to grow and to lose their spherical shape. After four hours of solar irradiation, the largest observed particle has a diameter of 638 nm and is still surrounded by small nanocrystals. The average particle diameter $\langle d\rangle$ can be approximately described by the growth law (\ref{eq:AverageGrowthLaw}), with initial average diameter $\langle d_0\rangle \approx 6.429$ and kinetic coefficient $k_\phi \approx 0.881$, where $d$ is in nanometers and $t$ is in minutes. While the particle size distribution initially stays log-normal, it eventually starts to deviate from the log-normal distribution. The resulting distribution does not fit the steady-state distributions predicted by the Ostwald ripening model either, which might be explained by the system being in a transient state.

\section*{Acknowledgements}

\noindent We are grateful to Jack Bonsak for making the colloidal silver particles, to Long-Qing Chen and Guang Sheng for providing numerical values of the kinetic coefficients, and to Chang Chuan You for supplying the temperature measurements with time. 


\end{document}